\documentclass{article}

\usepackage{color}

\usepackage{siunitx}
\usepackage{float}
\usepackage{booktabs}
\usepackage{standalone}
\usepackage{tikz}
\usetikzlibrary{shapes, arrows, positioning}
\usepackage{csquotes}
\usepackage{dsfont}

\usepackage{authblk}
\usepackage[round]{natbib}
\usepackage{amsmath}
\usepackage{amssymb}
\usepackage{mathtools}
\usepackage{bm}
\usepackage{url}
\usepackage[a4paper, margin=1in]{geometry}

\newcommand{\cC}{\mathcal{C}}
\newcommand{\R}{\mathbb{R}}

\title{Bivariate Postprocessing of Wind Vectors}

\author[1]{Ferdinand Buchner}
\author[2]{David Jobst}
\author[3]{Annette Möller}
\author[1,4]{Claudia Czado}

\affil[1]{Technical University of Munich, Applied Mathematical Statistics, Boltzmannstra{\ss}e 3, 85748 Garching bei München, Germany}
\affil[2]{University of Hildesheim, Institute of Mathematics - IMMI, Samelsonplatz 1, 31141 Hildesheim, Germany}
\affil[3]{Bielefeld University, Faculty of Business Administration and Economics, Universitätsstraße 25, 33615 Bielefeld, Germany}
\affil[4]{Technical University of Munich, Munich Data Science Institute (MDSI), Walther-von-Dyck-Straße 10, 85748 Garching bei München, Germany}

\begin{document}

\maketitle

\begin{abstract}
To quantify the uncertainty in numerical weather prediction (NWP) forecasts, ensemble prediction systems are utilized. Although NWP forecasts continuously improve, they suffer from systematic bias and dispersion errors. To obtain well calibrated and sharp predictive probability distributions, statistical postprocessing methods are applied to NWP output. 
Recent developments focus on multivariate postprocessing models incorporating dependencies directly into the model. We introduce three novel bivariate postprocessing approaches, and analyze their performance for joint postprocessing of bivariate wind vector components for 60 stations in Germany. Bivariate vine copula based models, a bivariate gradient boosted version of ensemble model output statistics (EMOS), and a bivariate distributional regression network (DRN) are compared to bivariate EMOS. The case study indicates that the novel bivariate methods improve over the bivariate EMOS approaches. The bivariate DRN and the most flexible version of the bivariate vine copula approach exhibit the best performance in terms of verification scores and calibration. 

\end{abstract}


\section{Introduction} \label{sec:introdution}

The state-of-the-art models for weather prediction are \textit{numerical weather prediction (NWP)} models that describe the physics of the atmosphere through a system of partial differential equations. Uncertainties in models and forecasts are addressed using ensemble prediction systems (\citealp{Bauer&2015}). Forecast ensembles typically suffer from biases and dispersion errors, indicating that these uncertainties are not adequately represented \citep{HamillColucci1997}. Hence, ensemble forecasts require a correction to be more accurate and reliable. 
For this task, statistical ensemble postprocessing models have been established to correct the NWP forecasts \citep{gneiting2014probabilistic}.

Early postprocessing (PP) models were typically univariate, i.e.\ they are applied to postprocess a single variable at a single station and lead time, thus ignoring potential dependencies. In recent years, researchers started to acknowledge the importance of incorporating multivariate dependencies explicitly into postprocessing models to obtain probabilistic forecasts that are physically more realistic. Multivariate dependencies include all sorts of physical relationships such as spatial or temporal patterns, dependence between lead times, and physical relationships between weather variables. The raw NWP ensemble forecasts naturally contain these physical dependencies, as they originate from a physical model describing the dynamics of the atmosphere. However, this inherent dependence structure is lost when a univariate postprocessing model is applied to the ensemble forecasts \citep{Schefzik&2013}. 
The most popular approach that uses the dependence information from ensemble forecasts is the Ensemble Copula Coupling (ECC, \citealp{Schefzik&2013}), which reorders samples from univariate predictive distributions according to the rank structure of the raw ensemble. Other researchers approached the issue of multivariate postprocessing from various perspectives, targeting different types of dependencies such as spatial, temporal, or inter-variable ones. 

To implicitly account for the circular nature of wind and to incorporate possible dependencies between wind speed and wind direction, a bivariate postprocessing model for the zonal and meridional ($u$- and $v$-) wind vector components is a reasonable approach. 
This analysis focuses on incorporating inter-variable dependencies between the $u$- and $v$-component explicitly into a postprocessing model. In recent years, serval researchers introduced postprocessing approaches for the $u$- and $v$-wind vector components. For example \cite{Pinson2012} utilizes a bivariate Gaussian distribution for the observed wind vector, however, generates a postprocessed forecast ensemble by estimating an ensemble specific dilation and translation factor based on the empirical correlation structure of the raw ensemble. The approach can be viewed as a variant of ECC \citep{Schefzik&2013}.
\cite{schuhen2012ensemble} introduce a postprocessing approach for the wind vector components based on a fully specified bivariate Gaussian distribution, rather than obtaining individually postprocessed ensemble members. As the correlation between the wind vector components mainly depends on the wind direction according to their analysis the correlation is modeled as a trigonometric function of the mean of the ensemble wind direction forecasts separately for different wind sectors. 
\cite{Lang&2019} generalize the bivariate Gaussian approach within the framework of distributional regression, where location, scale and correlation parameter are flexibly linked to covariates. They employ cyclic regression splines as model for the distribution parameters and estimate the parameters in a Bayesian framework.

More general methods of modeling inter-variable dependencies have also been proposed. \cite{moller2013multivariate} introduced a Gaussian copula approach (GCA) for joint postprocessing of an arbitrary number of weather variables. Later, GCA was applied for bivariate postprocessing of temperature and wind speed \citep{BaranMoeller2015, BaranMoeller2017}.

In GCA, each variable is postprocessed individually with a univariate approach, and the joint correlation structure is then added within the Gaussian copula framework.
This approach was quite successful due to its straightforward and simple estimation and is utilized as a benchmark method in several studies (see, e.g., \citealp{Lerch&2020, Chen&2024}). 
However, it lacks flexibility, specifically when it comes to high dimensional applications such as spatio-temporal settings, dependence modeling between a larger set of weather variables, or the inclusion of additional covariates into the model.
To introduce more flexibility and facilitate a more data-driven estimation and variable selection, \cite{moller2018vine} introduced a vine copula based PP approach based on D-vine regression by \cite{kraus2017d} for postprocessing temperature. 
\cite{jobst2023d} refined the D-vine regression for postprocessing wind speed, and made use of the forward variable selection algorithm by \cite{kraus2017d}. 
However, D-vine copula based PP approaches are univariate in nature.
Recently, the D-vine based postprocessing was extended to allow for temporal or spatial dependencies by parameterizing the dependence parameter associated with the bivariate copulas in the D-vine copula in terms of a generalized additive model (GAM, \citealp{jobst2025}) or a generalized linear model (GLM, \citealp{Jobst&2024b}). 
The recently introduced bivariate Y-vine structure \citep{tepegjozova2023bivariate} allows for estimating a genuine bivariate predictive distribution and is thus suitable for joint postprocessing of two weather quantities such as the wind vector components. The Y-vine based regression will be utilized in the following to obtain a novel bivariate postprocessing model. 

The remainder of the article is organized as follows: Section \ref{sec:data} provides an overview of the data. Section \ref{sec:methods} introduces the bivariate models compared in our case study. Section \ref{sec:verification} presents a short review of the verification methods used to evaluate the bivariate models. The results of the case study are presented in Section \ref{sec:results}. We close with a short conclusion and outlook in Section \ref{sec:outlook}.
\vspace*{-0.4cm}

\section{Data} \label{sec:data}

\begin{figure}
    \begin{center}
		\includegraphics[width = 0.5\linewidth]{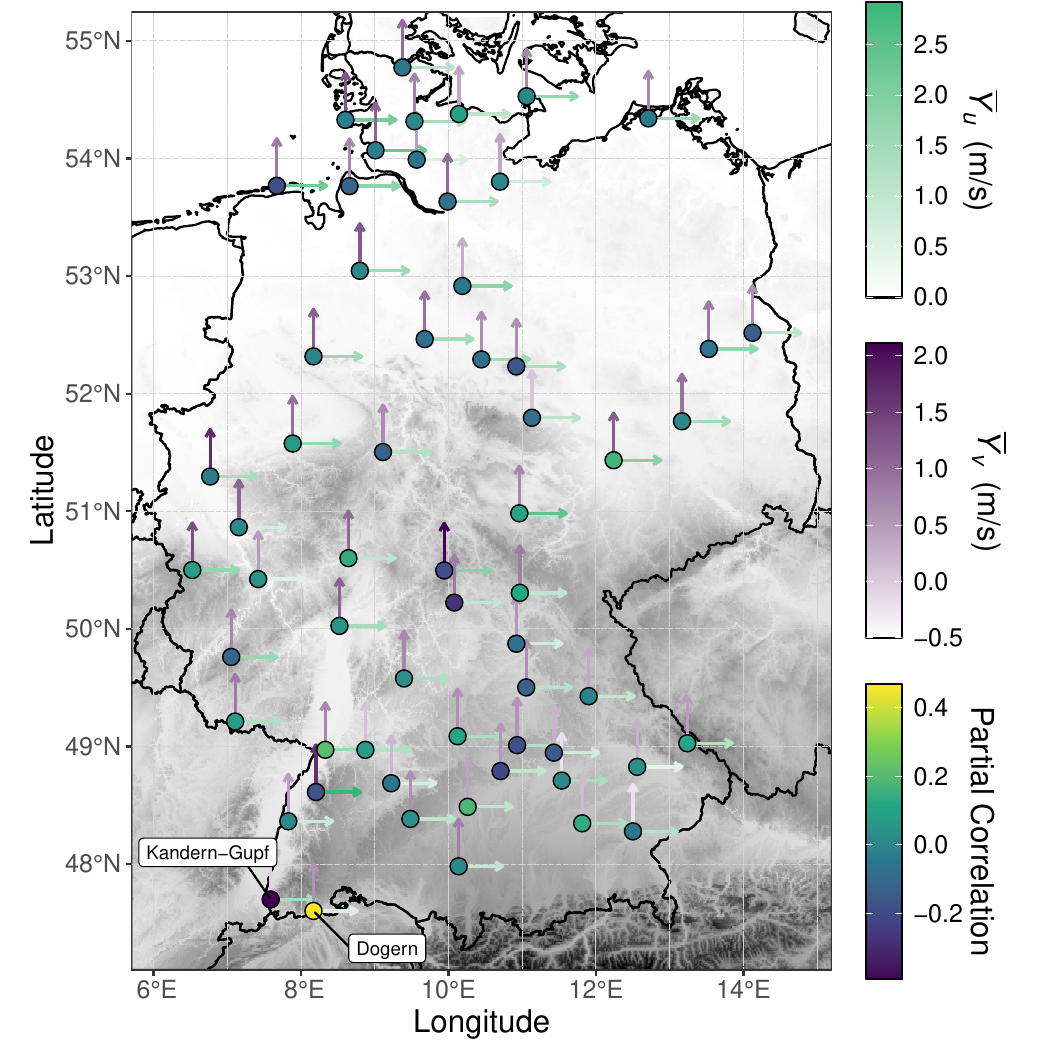}
    \end{center}	
    \caption{Observation stations with $10$ m surface $u$- and $v$-component averaged over all 880 training days.}
    \label{fig: stations}
\end{figure}

\paragraph{Observation and forecast data.} 
The goal is to jointly postprocess the $10$ m surface wind vector $u$- and $v$-component, corresponding to the zonal and meridional directions, at 1200 UTC for a forecast lead time of $24$ hours in Germany. The observations for the $10$ m surface wind vector $u$- and $v$-component are obtained from \cite{dwd} for each of the 60 observation stations marked by a point in Figure \ref{fig: stations}. The forecast ensemble, comprising 50 members of the perturbed ensemble along with a control forecast for various meteorological variables, is supplied by the \cite{ECMWF2021}. Since the forecast ensemble is initialized on a grid with resolution $0.25^\circ \times 0.25^\circ$, the forecasts are bilinearly interpolated from the four nearest grid points to the station using the {\sffamily R} package \textbf{akima} \citep{Akima1998}. The observation and forecast data are available daily from January 2, 2016 to December 31, 2020 (1824 days), excluding leap year days. \vspace*{-0.5cm}

\paragraph{Model training, validation and testing.} 
All methods, except for a neural network based method, are estimated locally, whereby a separate model is fitted for each station. For model training, we use the data spanning from January 2, 2016 to May 31, 2018 (880 days) as a fixed static training period. However, some of the considered methods also employ parts of the training data as validation data for hyperparameter tuning. Eventually, all methods are evaluated on the test data from June 1, 2018 to December 31, 2020 (944 days). \vspace*{-0.5cm}

\paragraph{Notation.}
\begin{table}
    \centering
    \resizebox{0.75\linewidth}{!}{
    \begin{tabular}{ll} 
        \toprule
        Weather variable ($q$) & Description \\ 
        \midrule
        $u$ & $10$ m surface wind vector $u$-component \\
        $v$ & $10$ m surface wind vector $v$-component \\
        $wspd$ & $10$ m surface wind speed \\
        $wdir$ & $10$ m surface wind direction \\
        $temp$ & $2$ m surface temperature \\
        $pres$ & surface pressure \\
        $sh$ & Logarithmic specific humidity at 850 hPa \\
        \bottomrule
    \end{tabular}
    }
    \caption{Set of considered weather variables.}
    \label{tab::variables}
\end{table}

For a variable $q$ in Table \ref{tab::variables} with 50 ensemble forecasts $X_{q}^{1}, \dots, X_{q}^{50}$, we define the ensemble mean $\overline{X}_{q}$ and the $\log$-transformed ensemble standard deviation $X_{q}^{sd}$ by
\begin{align*}
    \overline{X}_{q} \coloneqq \frac1{50} \sum_{i = 1}^{50} X_{q}^{i} \quad \text{and} \quad X_{q}^{sd} \coloneqq \ln\sqrt{\frac1{49} \sum_{i = 1}^{50} {(X_{q}^{i} - \overline{X}_{q})}^2}, \label{eq: ensemble_mean}
\end{align*}
and the control forecast by $X_{q}^{ctrl}$. The ensemble mean $\overline{X}_{wdir}$ of the $10m$ surface wind direction is calculated as
\begin{equation} \label{eq: wdir}
    \overline{X}_{wdir} = \frac{180 \cdot \mathrm{arctan2}(\overline{X}_{v}, \overline{X}_{u})}{\pi} \mod 360.
\end{equation}
Similarly, the $10m$ surface wind direction $X_{wdir}^{ctrl}$ is obtained by plugging in $X_u^{ctrl}$ and $X_v^{ctrl}$ into \eqref{eq: wdir} instead. We define two different sets of covariates 
\begin{small}
\begin{align*}
    \cC & \coloneqq \left\{ \overline{X}_{q}, X_{q}^{ctrl}, X_{q}^{sd} \, | \, q \in \{ u, v, wspd \}\right\} \cup \left\{ \overline{X}_{wdir}, X_{wdir}^{ctrl} \right\}, \\
    \cC^+ & \coloneqq \left\{ \overline{X}_{q}, X_{q}^{ctrl}, X_{q}^{sd} \, | \, q \in \{ u, v, temp, pres, sh \} \right\}, 
\end{align*}
\end{small}
used for the subsequently presented methods. The set $\cC$ contains $11$ variables and $\cC^+$ contains $15$ variables.

\section{Methods} \label{sec:methods}

For the case study in Section \ref{sec:results} we compare existing bivariate methods to completely novel bivariate approaches. In this section, the models are described in more detail. Table \ref{tab:overview_models} provides an overview of all considered models. 
In the following, let $\bm{Y} \coloneq (Y_u,Y_v)\in \R^2$ be the response vector of the observed $u$- and $v$-wind components, and let $\bm{X} \coloneq (X_1,\ldots, X_p) \in \R^p$ be a vector of covariates.

\begin{table}
    \centering
    \resizebox{0.75\linewidth}{!}{
    \begin{tabular}{ll}
   \toprule
     Abbreviation  & Description  \\ \midrule
     \textbf{IND-EMOS} & Independent EMOS for each response  \\
     \textbf{BIV-EMOS} & Bivariate EMOS with correlation structure \\
     \textbf{BIV-EMOS-GB} & Gradient-boosted bivariate EMOS \\
     \textbf{BIV-DRN} & Bivariate Distributional Regression Network \\
     \textbf{BIV-YV-G} & Bivariate Y-vine, Gaussian pair copulas (PCs) \\
     \textbf{BIV-YV-P} & Bivariate Y-vine, parametric pair copulas (PCs) \\
     \textbf{BIV-YV-ALL} & Bivariate Y-vine,  parametric and nonparametric PCs \\ 
     \bottomrule
    \end{tabular}
    }
    \caption{Considered models and their abbreviation.}
    \label{tab:overview_models}
\end{table}

\subsection{Bivariate Ensemble Model Output Statistics}
\label{sec: Bivariate Ensemble Model Output Statistics}

The state-of-the-art approach for univariate postprocessing is ensemble model output statistics (EMOS, \citealp{gneiting2005calibrated}). Following \cite{schuhen2012ensemble}, the bivariate EMOS models and our proposed extensions assume a conditional bivariate normal distribution $\bm{Y} \vert \bm{X}=\bm{x} \sim \mathcal{N}_2(\bm{\mu}(\bm{x}),\bm{\Sigma}(\bm{x}))$ for the wind vector, where the predictors are linked to 
its location, scale, and correlation parameters.
Its conditional probability density function evaluated at $\bm{y}:=(y_u,y_v)\in \R^2$ is given by 
\begin{align}
    &f_{Y_u,Y_v\vert \bm{X}}(\bm{y} \vert \bm{\mu}(\bm{x}), \bm{\Sigma}(\bm{x})) 
     :=\frac{1}{2\pi \sqrt{\mathrm{det}(\bm{\Sigma}(\bm{x}))}} \cdot \exp\left(-\frac{1}{2}(\bm{y}-\bm{\mu}(\bm{x}))\bm{\Sigma}^{-1}(\bm{x})(\bm{y}-\bm{\mu}(\bm{x}))^T\right),\label{eq: biv_normal}
\end{align}
with conditional mean vector $\bm{\mu}(\bm{x}):=(\mu_u(\bm{x}),\mu_v(\bm{x}))\in \R^2$ and conditional covariance matrix $\bm{\Sigma}(\bm{x})$ with marginal variances $\sigma_u^2(\bm{x}), \sigma_v^2(\bm{x}) \in (0, \infty)$ and correlation parameter $\rho(\bm{x}) \in (-1, 1)$.
The distribution parameters are connected to linear predictors $\eta_{\bullet}$, $\bullet\in \{\mu_u, \mu_v, \sigma_u, \sigma_v, \rho\}$, using the following link functions
\begin{align}
    &\mu_u(\bm{x})=\eta_{\mu_u}(\bm{x}),\,\, \mu_v(\bm{x})=\eta_{\mu_v}(\bm{x}),\,\, \ln(\sigma_u(\bm{x})) =\eta_{\sigma_u}(\bm{x}), \nonumber\\ &\ln(\sigma_v(\bm{x}))=\eta_{\sigma_v}(\bm{x}),\,\,  
\frac{\rho(\bm{x})}{\sqrt{1-\rho(\bm{x})^2}}=\eta_{\rho}(\bm{x}).
\label{eq: link_funs}
\end{align} 
The linear predictors are defined in terms of one or more covariates, depending on the specific model setting. For the bivariate EMOS models, exact definitions of the respective linear predictors are given in \eqref{eq:linpredsEmosmu} and \eqref{eq:linpredsEmossigma}.

For joint postprocessing of the $u$- and $v$-wind component, a modification of the bivariate EMOS version of \cite{Lang&2019} is utilized. As benchmark models for our case study in Section \ref{sec:results} we introduce two variants of the so-called baseline model (BLM) analyzed in \cite{Lang&2019}. As first benchmark model we study an adapted version of the so-called BLM-0 model in \cite{Lang&2019}, called \textbf{IND-EMOS} in our analysis. \textbf{IND-EMOS} combines two univariate regression models that postprocess the $u$- and $v$-component separately with zero correlation, i.e. the wind components are assumed to be conditionally independent. Consequently, the location and scale parameter in the $Y_u$ model uses only summary statistics of the ensemble forecasts for $Y_u$, while the model for $Y_v$ only uses summary statistics of the ensemble forecasts for $Y_v$.

The second benchmark model in our case study is a combination of a BLM for location and scale parameter with the so-called ADV correlation structure for the correlation parameter, utilized in the rotation allowing Model (RAM) in \cite{Lang&2019}. The ADV correlation structure uses summary statistics of the wind direction and wind speed as covariates, and additionally models a linear interaction between these two covariates. 
BLM-ADV therefore results from combining the original BLM and RAM-ADV introduced in \cite{Lang&2019}. We subsequently call it \textbf{BIV-EMOS} and utilize the predictor variables in the set $\cC$ defined in Section \ref{sec:data} for the model. 
In order to link the predictor variables from the set $\cC$ to the distribution parameters in \textbf{IND-EMOS} and \textbf{BIV-EMOS}, we employ the link functions as in \eqref{eq: link_funs}. 
However, contrary to \cite{Lang&2019}, we utilize linear predictors instead of additive predictors made up of cubic smooth functions, to be directly comparable to the setup used in \textbf{BIV-EMOS-GB} introduced in Section \ref{sec: Gradient-Boosted Bivariate Ensemble Model Output Statistics}. 

In particular, for location and scale parameters, we use
\begin{align} 
\eta_{\mu_{\bullet}}(\bm{x})    &= \alpha_{0,\mu_{\bullet}} + \alpha_{1,\mu_{\bullet}} \overline{x}_{\bullet} + \alpha_{2,\mu_{\bullet}} x^{ctrl}_{\bullet}, \label{eq:linpredsEmosmu} \\
\eta_{\sigma_{\bullet}}(\bm{x}) &= \alpha_{0,\sigma_{\bullet}} + \alpha_{1,\sigma_{\bullet}} x^{sd}_{\bullet}, \label{eq:linpredsEmossigma}
\end{align}
with coefficients $\alpha_{0,\bullet}, \alpha_{1,\bullet}, \alpha_{2,\bullet} \in \R$, $\bullet \in \{u,v\}$.

For \textbf{IND-EMOS} the correlation parameter is assumed to be $\rho(\bm{x})=0$, i.e. the covariance matrix $\boldsymbol\Sigma(\bm{x})$ is a diagonal matrix with $\sigma_u^2(\bm{x})$ and $\sigma_v^2(\bm{x})$ on the diagonal. 
For \textbf{BIV-EMOS} the predictor variables are linked to the correlation parameter as in the ADV-correlation model:
\begin{align}
\eta_{\rho}(\bm{x}) &= \alpha_{0,\rho}  + \alpha_{1,\rho} \overline{x}_{wdir} \overline{x}_{wspd} + \alpha_{2,\rho} x^{ctrl}_{wdir} x^{ctrl}_{wspd},
\end{align}
with coefficients $\alpha_{0,\rho}, \alpha_{1,\rho}, \alpha_{2,\rho} \in \R$. Both bivariate EMOS variants are estimated with the {\sffamily R} package \textbf{bamlss} \citep{Umlauf&2018} within a Bayesian framework via Markov chain Monte Carlo samples as in \cite{Lang&2019}. 

\subsection{Gradient-Boosted Bivariate Ensemble Model Output Statistics} \label{sec: Gradient-Boosted Bivariate Ensemble Model Output Statistics}

With the increasing number of potential covariates as e.g. in the set $\mathcal{C}^+$, the estimation procedure for distributional regression models, such as EMOS, needs to be reconsidered in order to avoid issues like multicollinearity and overfitting. 
To address these challenges, a possible approach is to employ a regularization technique for model estimation, such as gradient-based boosting, which was originally proposed by \cite{Friedman2000, Friedman2001}. This approach was later extended to the framework of distributional regression, allowing simultaneous estimation and variable selection for all parameters of a univariate distribution \citep{Mayr2012}. Recent work has further adapted gradient boosting to multivariate distributions, as shown by \cite{Stroemer2023}, who examined various bivariate distributions in simulations and biomedical applications. In the context of ensemble postprocessing, \cite{Simon2017} have already considered a truly multivariate Gaussian distribution for the joint correction of ensemble forecasts for different lead times in a gradient-boosting setting. Recently, \cite{Jobst&2024b} suggested a gradient-boosted estimation of a conditional vine copula for the same application context.

As gradient boosting-based estimation of multivariate distributional regression models have demonstrated considerable suitability and yielded promising outcomes in the context of ensemble postprocessing \citep{Simon2017, Jobst&2024b}, we also adopt this estimation technique for EMOS in our application to the bivariate postprocessing of the $u-$ and $v-$  wind vector components.

In order to facilitate comparability among the studied approaches, to enhance computational efficiency, and to enable the quantification of variable importance \citep{Messner&2017}, we initially standardize the response variables and the covariates $\xi \in  \mathcal{C}^+$
\begin{align}
    Z_u:=\frac{Y_u-\widehat{\mu}_{u}}{\widehat{\sigma}_{u}},\quad Z_v:=\frac{Y_v-\widehat{\mu}_{v}}{\widehat{\sigma}_{v}},\quad S_\xi:=\frac{\xi-\widehat{\mu}_\xi}{\widehat{\sigma}_\xi},\label{eq: standardized_covariates} 
\end{align}
by its respective sample mean $\widehat{\mu}_{\xi}$ and sample standard deviation $\widehat{\sigma}_{\xi}$, where $\xi \in \mathcal{C}^+$.
We are interested in the conditional distribution $\bm{Z}\vert \bm{S}=\bm{s}\sim \mathcal{N}_2(\bm{\mu}_{\bm{Z}}(\bm{s}), \bm{\Sigma}_{\bm{Z}}(\bm{s}))$ with density as in \eqref{eq: biv_normal}, where $\bm{Z}:=(Z_u,Z_v)$ denotes the standardized response and $\bm{S}:=(S_1,\ldots, S_{15})\in \R^{15}$ the standardized covariates. The conditional means are denoted by $\mu_{Z_u}(\bm{s})$ and $\mu_{Z_v}(\bm{s})$, while the conditional marginal variances are given by $\sigma_{Z_u}(\bm{s})^2:=\mathrm{Var}(Z_u\,\vert\, \bm{S}=\bm{s})$, $\sigma_{Z_v}(\bm{s})^2:=\mathrm{Var}(Z_v\,\vert\, \bm{S}=\bm{s})$ and the conditional correlation by  $\rho(\bm{s}):=\mathrm{Cor}(Z_u,Z_v\,\vert\, \bm{S}=\bm{s})$. 
All $p=15$ standardized covariates from $\mathcal{C}^+$ are allowed for each of the five linear predictors in \eqref{eq: link_funs} as 
$\eta_{\bullet}(\bm{s}):=\alpha_{0,\bullet}+\alpha_{1,\bullet}s_1+\ldots+\alpha_{15, \bullet}s_{15}$
with coefficients $\alpha_{0,\bullet},\ldots, \alpha_{15,\bullet}\in \R$ for all $\bullet\in \{\mu_{Z_u}, \mu_{Z_v},$ $\sigma_{Z_u}, \sigma_{Z_v}, \rho\}$. \\
We use the non-cyclic gradient-boosting estimation algorithm \citep{Thomas2017, Messner&2017}, which updates in each iteration only one regression coefficient across all linear predictors, i.e. the one which yields to the highest loss reduction in terms of the negative log-likelihood. Since all regression coefficients have been initially set to zero this iterative procedure incrementally adjusts the coefficients away from zero. Due to early stopping, informative covariates will receive coefficients different from zero and are thus selected for the model, while coefficients of non-informative covariates remain zero and are thus ignored. To avoid overfitting, we consider a learning rate (step length) of $\nu=0.1$ and employ 10-fold cross-validation as an early stopping criterion to determine the optimal number of boosting iterations, subsequent to an initial phase of 2000 boosting iterations. For the majority of fitted models, an optimal number of boosting iterations between $150$ and $500$ iterations has been selected.

After the distributional regression model has been estimated via gradient-boosting, the parameters of the conditional distribution on the original scale $\bm{Y} \vert \bm{X}=\bm{x} \sim \mathcal{N}_2(\bm{\mu}(\bm{x}),\bm{\Sigma}(\bm{x}))$ can be recovered via 

{\small
\begin{align}
\bm{\mu}(\bm{x})&=(\mu_{Z_u}(\bm{s})\widehat{\sigma}_{u}+\widehat{\mu}_{u}, \mu_{Z_v}(\bm{s})\widehat{\sigma}_{v}+\widehat{\mu}_{v}),\\ 
\bm{\Sigma}(\bm{x})&=
\begin{pmatrix}
(\sigma_{Z_u}(\bm{s})\widehat{\sigma}_{u})^2 & \rho(\bm{s})\sigma_{Z_u}(\bm{s})\widehat{\sigma}_{u}\sigma_{Z_v}(\bm{s})\widehat{\sigma}_{v}\\
\rho(\bm{s})\sigma_{Z_u}(\bm{s})\widehat{\sigma}_{u}\sigma_{Z_v}(\bm{s})\widehat{\sigma}_{v} & 
(\sigma_{Z_v}(\bm{s})\widehat{\sigma}_{v})^2
\end{pmatrix}.
\label{eq: mu_sigma_obsscale}
\end{align}
}
This gradient-boosting extension of bivariate EMOS is referred to as \textbf{BIV-EMOS-GB}. Model estimation is performed by the {\sffamily R} package \textbf{gamboostLSS} \citep{Hofner2016}. 

\subsection{Bivariate Distributional Regression Neural Network}
\label{sec: Bivariate Distributional Regression Neural Network}

Instead of estimating the distribution parameters of the bivariate normal distribution via gradient-boosting, we can estimate them based on neural networks. Neural network-based ensemble postprocessing methods have shown to greatly outperform state-of-the-art postprocessing for univariate responses such as surface temperature \citep{Rasp2018} or wind gust \citep{Schulz2021a}. We extend the distributional regression neural network framework of \cite{Rasp2018} from a univariate to a multivariate setting, thereby providing a novel contribution.

Similar to \textbf{BIV-EMOS-GB}, we consider the bivariate normal distribution as predictive distribution, but we do not estimate a separate distributional regression model for each station. Instead, we employ an technique called \textit{embedding} \citep{Guo2016} 
arising from the field of natural language processing and recommender systems. In a nutshell, an embedding is a mapping
\begin{align*}
M:\{1,2,\ldots,60\} \to I^{n_{\mathrm{emb}}},\quad l\mapsto \bm{E}=(E_1, \ldots, E_{n_{\mathrm{emb}}}),
\end{align*}
which maps in our specific case each of the 60 stations, characterized by the station identity (ID), to a vector of $n_{\mathrm{emb}}$ latent features. These latent features are given in terms of $E_k\sim \mathrm{Uniform}(I)$ with $I:=[-0.05,0.05]$ for $k=1,\ldots, n_{\mathrm{emb}}$, where the size $n_{\mathrm{emb}}$ of the embedding vectors needs to be tuned.
Consequently, the station embedding associates each station ID with a corresponding embedding vector, thereby enabling the neural network model to learn station-specific characteristics.

\begin{figure}
  \centering
  \includegraphics[width = 0.5\linewidth]{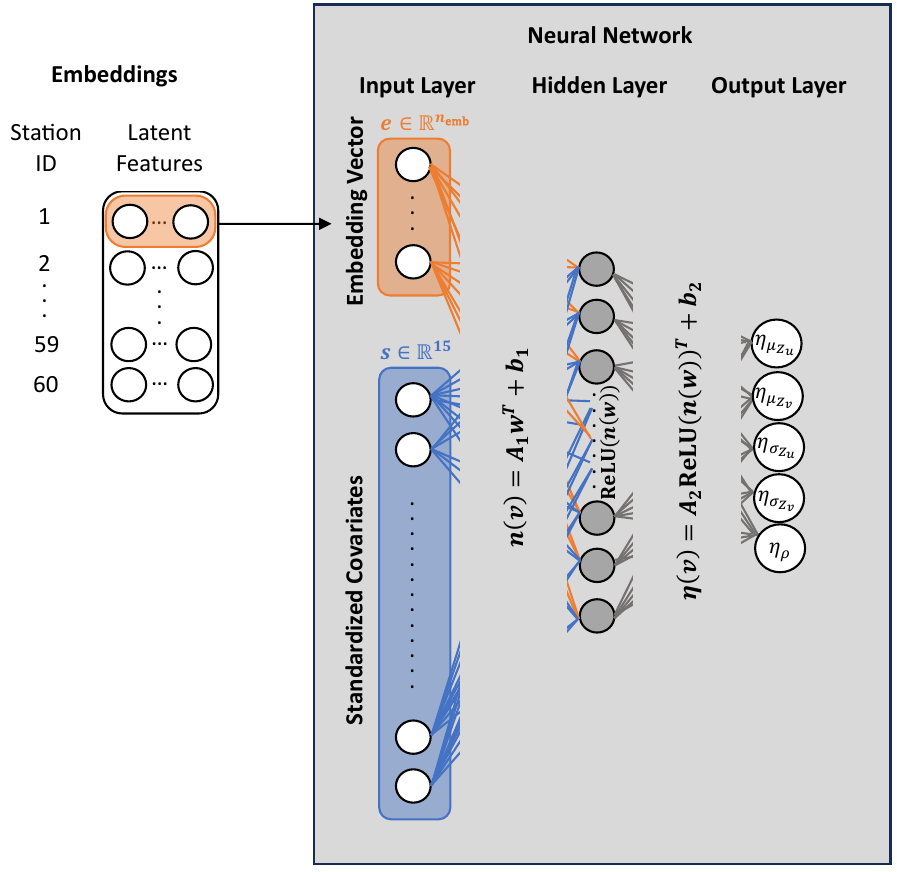}
  \caption{Schematic neural network with one hidden layer adapted from \cite{Rasp2018}.}
  \label{fig: nnet}
\end{figure}

Besides the embedding vector $\bm{E}=(E_1,\ldots, E_{n_{\mathrm{emb}}})$, the 15 standardized covariates $\bm{S}=(S_1,\ldots, S_{15})$ from $\mathcal{C}^+$ of all stations are provided for the input layer of the neural network. Not only the covariates but also the responses are standardized by its sample mean and sample standard deviation as introduced in Equation \eqref{eq: standardized_covariates}. Consequently, the variables $\bm{W}:=(\bm{E},\bm{S})\in \R^{d}$ with $d:=n_{\mathrm{emb}}+15$ are considered as covariates and its realizations $\bm{w}\in \R^{d}$ are used in the input layer for each station and time point. Eventually, we estimate the parameters of the  conditional distribution $\bm{Z}\vert \bm{W}=\bm{w}\sim \mathcal{N}_2(\bm{\mu}_{\bm{Z}}(\bm{w}), \bm{\Sigma}_{\bm{Z}}(\bm{w}))$ based on the distributional neural network, whose architecture is illustrated in Figure \ref{fig: nnet} and briefly summarized below. 

Similar to \cite{Rasp2018}, initial tests have shown no meaningful improvements when more than one hidden layers are used in the neural network. Therefore, we employ only one hidden layer in order to keep our neural network as simple and parsimonious as possible. This hidden layer consists of $n$ nodes (neurons), whose values are obtained from the linear mapping $\bm{n}(\bm{w}):=\bm{A}_1\bm{w}^T+\bm{b}_1$, where $\bm{A}_1\in \R^{n \times d}$ and $\bm{b}_1\in \R^n$ denote the weight matrix and the bias term of the hidden layer, respectively. In addition, the node values $\bm{n}(\bm{w})\in \R^n$ of the hidden layer are passed through a so-called activation function, which allows to learn non-linear relationships. We choose the rectified linear unit (ReLU, \citealp{Householder1941}) as activation function. Afterwards, the values for the output layer, which consists in our case of the five regression predictors $\bm{\eta}:=(\eta_{\mu_{Z_u}}, \eta_{\mu_{Z_v}}, \eta_{\sigma_{Z_u}}, \eta_{\sigma_{Z_v}}, \eta_{\rho})\in \R^5$ defining the conditional distribution $\mathcal{N}_2(\bm{\mu}_{\bm{Z}}(\bm{w}), \bm{\Sigma}_{\bm{Z}}(\bm{w}))$, are obtained as follows
\begin{align*}
    \bm{\eta}(\bm{w}):=\bm{A}_2\mathrm{ReLU}(\bm{n}(\bm{w}))^T+\bm{b}_2,
\end{align*}
where $\bm{A}_2\in \R^{5\times n}$ and $\bm{b}_2\in \R^{5}$ denote the weight matrix and the bias term of the output layer, respectively. Eventually, the regression predictors $\eta_{\mu_{Z_u}}, \eta_{\mu_{Z_v}}, \eta_{\sigma_{Z_u}}, \eta_{\sigma_{Z_v}}, \eta_{\rho}$ are mapped using the link functions in \eqref{eq: link_funs} to the parameters of the bivariate normal distribution. Since we have employed standardized variables, we re-obtain the parameters of $\bm{Y} \vert \bm{X}=\bm{x} \sim \mathcal{N}_2(\bm{\mu}(\bm{x}),\bm{\Sigma}(\bm{x}))$ analogously to \eqref{eq: mu_sigma_obsscale}.
Regarding model estimation, both weight matrices and bias terms are learned by minimizing the negative log-likelihood of the conditional bivariate normal distribution in the training data using the stochastic gradient descent (SGD) based on the adaptive moment estimation (Adam, \citealp{Kingma2014}) 
algorithm. In order to account for the inherent uncertainty in the training due to the SGD, we fit an ensemble of ten neural networks. The estimated distribution parameters of these ten neural networks are averaged in a final step. For model estimation, the {\sffamily R} packages \textbf{keras3} \citep{Kalinowski2024} and  \textbf{tensorflow} \citep{Allaire2017} have been used. This approach is called \textit{bivariate distributional regression neural network}, abbreviated as \textbf{BIV-DRN}, in the following. 

\subsection{Bivariate Vine Copula Regression} \label{sec: yvine}
\begin{figure}
    \centering
    \includegraphics[width = 0.75\linewidth]{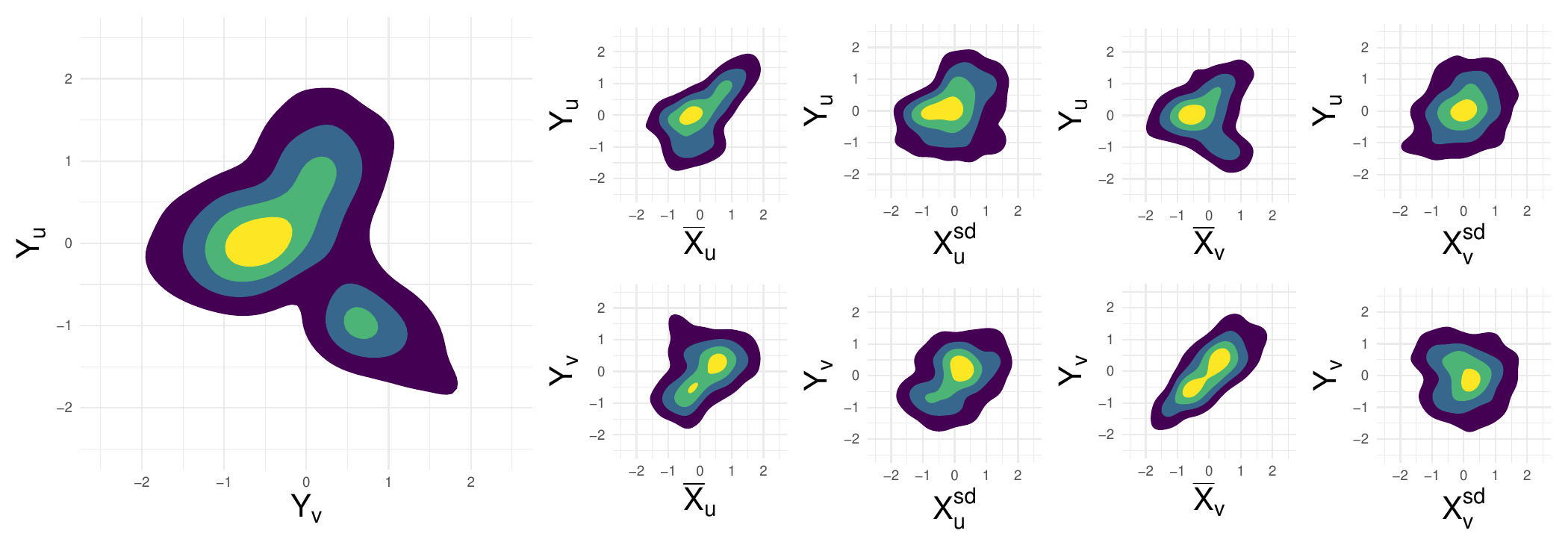}
    \caption{Empirical marginally normalized contour plots for observed and forecast wind vector component variables at station Kandern-Gupf for training data. Non-elliptical contours imply non-Gaussian dependence.}
    \label{fig: enc_plot}
\end{figure}

So far, all models presented assume the forecast distribution to be the bivariate normal distribution. This implies that the marginal distributions of $Y_u$ and $Y_v$ are univariate Normal, and that the dependence between both responses is Gaussian. Both implications might not be justified by the data for all stations, see Figure~\ref{fig: enc_plot}. We follow a copula approach that allows to model marginal distributions and dependence structure between variables separately from each other \citep{sklar1959fonctions}. Copulas have already been used in the context of postprocessing. Two prominent non-parametric examples are the empirical copula coupling \citep{Schefzik&2013} and Schaake shuffle \citep{Clark&2004, Schefzik2016b} which are both based on the empirical copula. \cite{moller2013multivariate} use a multivariate Gaussian copula to reestablish the dependence structure of several univariate postprocessed variables. A general overview can be found in \cite{SchefzikMoeller2018}.

\begin{figure}
    \centering
    \includegraphics[width = 0.75\linewidth]{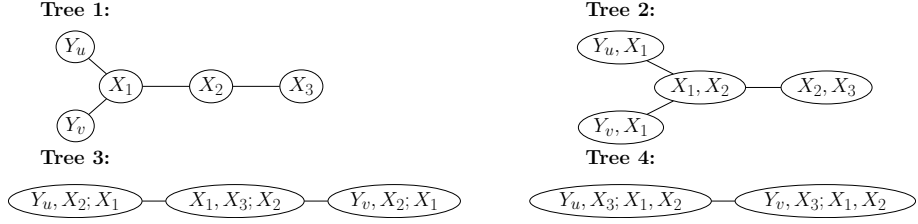}
    \caption{Y-vine tree structure with 3 covariates.}
    \label{fig:yvinetree}
\end{figure}

While the result in \cite{sklar1959fonctions} allows to separate modeling of the margins and the dependence between them, parametric copula approaches often assume the copula to be of a specific family, i.e.\ the multivariate Gaussian copula \citep{moller2013multivariate}. However, the dependence between pairs of variables can vary. Vine copulas \citep{bedford2001probability} avoid this problem by constructing multivariate distributions out of bivariate copulas (pair copula) building blocks using conditioning \citep{Joe96}. These pair copulas are structured into several linked trees which together form a vine copula. A more detailed introduction to vine copulas can be found in \cite{czado2019} and \cite{joe-2014}. \cite{czado2022vine} give a review of recent vine copula developments. \cite{moller2018vine} and \cite{jobst2023d, jobst2025} use the class of D-vine copulas for univariate postprocessing, and \cite{Jobst&2024b} for multivariate postprocessing of several lead times.

To estimate the bivariate conditional distribution of $(Y_u, Y_v) | \bm X = \bm x$, we employ the bivariate Y-vine copula regression model  introduced by \cite{tepegjozova2023bivariate}. For this, the joint distribution of $(Y_u, Y_v, \bm X)$ is written as 
\begin{align*}
    & F_{Y_u, Y_v, \bm X} (y_u, y_v, \bm x) \\ 
    & = C_{V_u, V_v, \bm U} (F_{Y_u}(y_u), F_{Y_v}(y_v), F_{X_1}(x_1), \dots, F_{X_p}(x_p)),
\end{align*}
conditioning on the covariate vector $\bm X$ then yields the desired forecast distribution
\begin{align*}
    & F_{Y_u, Y_v | \bm X}(y_u, y_v | \bm x) \\
    & = C_{V_u, V_v | \bm U}(F_{Y_u}(y_u), F_{Y_v}(y_v) | F_{X_1}(x_1), \dots, F_{X_p}(x_p)).
\end{align*}
Here, $V_u = F_{Y_u}(Y_u)$, $V_v = F_{Y_v}(Y_v)$ and $U_i = F_{X_i}(X_i)$, $i = 1, \dots, p$, denote the probability integral transformed responses and covariates. Further set $\bm U = (U_1, \dots, U_p)$. The $p + 2$-dimensional copula $C_{V_u, V_v, \bm U}$ is constructed as a Y-vine copula and $C_{V_u, V_v | \bm U}$ denotes the bivariate conditional distribution of $(V_u, V_v) | \bm U = \bm u$, which arises from the copula $C_{V_u, V_v, \bm U}$. In this class of vine copulas, every tree has the shape of the letter \enquote{Y}, see Figure~\ref{fig:yvinetree}. It is suited for bivariate regression models, since the bivariate conditional density can be written as a product of pair copula densities without needing integration. The conditional density of $(Y_u, Y_v) | (X_1, X_2, X_3)$ following the vine tree structure given in Figure~\ref{fig:yvinetree} is
\begin{align} \label{eq: cond_yvine_pdf}
    & f_{Y_u, Y_v | X_1, X_2, X_3}(y_u, y_v | x_1, x_2, x_3) =  f_{Y_u}(y_u) \cdot f_{Y_v}(y_v) \nonumber \\
    & \cdot c_{V_u, U_1}(F_{Y_u}(y_u), F_{X_1}(x_1)) \cdot c_{V_v, U_1}(F_{Y_v}(y_v), F_{X_1}(x_1)) \nonumber \\
    & \cdot c_{V_u, U_2; U_1}(F_{Y_u | X_1}(y_u | x_1), F_{X_2 | X_1}(x_2 | x_1)) \nonumber \\
    & \cdot c_{V_v, U_2; U_1}(F_{Y_v | X_1}(y_v | x_1), F_{X_2 | X_1}(x_2 | x_1)) \nonumber \\
    & \cdot c_{V_u, U_3; U_1, U_2}(F_{Y_u | X_1, X_2}(y_u | x_1, x_2), F_{X_3 | X_1, X_2}(x_3 | x_1, x_2)) \nonumber \\
    & \cdot c_{V_v, U_3; U_1, U_2}(F_{Y_v | X_1, X_2}(y_v | x_1, x_2), F_{X_3 | X_1, X_2}(x_3 | x_1, x_2)) \nonumber \\
    & \cdot c_{V_u, V_v; U_1, U_2, U_3}(F_{Y_u | X_1, X_2, X_3}(y_u | x_1, x_2, x_3), \nonumber \\
    & \phantom{ \cdot c_{V_u, V_v; U_1, U_2, U_3}((} F_{Y_v | X_1, X_2, X_3}(y_v | x_1, x_2, x_3)).
\end{align}
Terms of the form $c_{V_u, U_2; U_1}(F_{Y_u | X_1}(y_u | x_1), F_{X_2 | X_1}(x_2 | x_1))$ are bivariate copula densities associated with $(Y_u, X_2)$ conditioned on $X_1$. The conditioning on the specific value of $X_1$ appears only in the margins. Hence, we make the simplifying assumption, which is typically used for estimation tractability in high dimensions. It states $c_{V_u, U_2; U_1}(\cdot, \cdot; x_1) = c_{V_u, U_2; U_1}(\cdot, \cdot)$. For a discussion, see \cite{nagler2025simplified}.

All arguments of copula densities in~\eqref{eq: cond_yvine_pdf} are obtained recursively from the Y-vine structure. There are $p! = 15! = 1.307674 \times 10^{12}$ possible orderings in which the covariates in $\cC^+$ can enter the Y-vine regression model. We employ a stepwise forward selection algorithm that automatically chooses the covariate that improves the model the most based on an adjusted conditional BIC in each step. It was first proposed by \cite{kraus2017d} for univariate D-vine regression models and later adapted for Y-vine regression by \cite{tepegjozova2023bivariate}.\\
Marginal distributions in~\eqref{eq: cond_yvine_pdf} are obtained using kernel density estimation and pair copulas can be chosen for each bivariate (conditional) copula individually. We fit three Y-vine regression models, one allowing only Gaussian pair copulas, one with parametric pair copulas, and one with parametric and non-parametric pair copulas. The models are denoted by \textbf{BIV-YV-G}, \textbf{BIV-YV-P}, and \textbf{BIV-YV-ALL}, respectively. For a list of implemented pair copula families see \cite{rvinecopulib2025}. These models are fitted using the {\sffamily R} package \textbf{bivinereg} \citep{bivinereg}.

\section{Verification Measures} \label{sec:verification}
The goal of probabilistic forecasting is to \enquote{maximize the sharpness of the forecast distribution $F$ subject to calibration} (\citealp{GneitingRaftery2007, gneiting2014probabilistic}). \textit{Calibration} refers to the statistical compatibility between the forecast distribution $F$ and the verifying observations. \textit{Sharpness} refers to the spread of the forecast distribution $F$ and is a property of the forecasts only. 

In our setting, the forecast distribution is the bivariate conditional distribution of $(Y_u, Y_v)$ given $\bm X = \bm x$, that is $F = F_{(Y_u, Y_v)| \bm X}(\cdot, \cdot | \bm x)$ (in the case of \textbf{BIV-DRN} we have $F = F_{(Y_u, Y_v)| \bm W}(\cdot, \cdot | \bm w)$ with $\bm W$ defined as in Section~\ref{sec: Bivariate Distributional Regression Neural Network}). Hence, our models need to be evaluated with methods designed for bivariate (or more general multivariate) distributions, see \cite{gneiting2008assessing}. In the following, we define the verification methods that we use for generic bivariate distributions $F$ with probability density function $f$. Note that in Section~\ref{sec:results}  the distribution $F$ and density $f$ are replaced by the estimated bivariate conditional distribution $\hat F_{Y_u, Y_v | \bm X}$ and the estimated bivariate conditional density $\hat f_{Y_u, Y_v | \bm X}$ of our fitted models, respectively. Let $\bm y = (y_1, y_2) \in \R^2$ be a verifying observation. To calculate the scores, we use the {\sffamily R} packages \textbf{eppverification} \citep{Jobst2021} and \textbf{forecastcalibration}, the latter is available at \url{https://github.com/FK83/forecastcalibration}.

\paragraph{Proper scoring rules.} 
Proper scoring rules \citep{GneitingRaftery2007} are used to simultaneously evaluate the sharpness and calibration of $F$ at $\bm y$. A commonly used proper scoring rule for multivariate forecast distributions is the \textit{energy score} ($\mathrm{ES}$, \citealp{GneitingRaftery2007}), which generalizes the univariate continuous ranked probability score (CRPS, \citealp{GneitingRaftery2007}). The ES is defined as
\begin{equation} \label{eq:es}
    \mathrm{ES}(F, \bm y) \coloneqq \mathds{E}_F \lVert \bm R - \bm y \rVert - \frac1{2} \mathds{E}_F \lVert \bm R - \bm R^* \rVert,
\end{equation}
where $\lVert\, \cdot\, \rVert$ denotes the Euclidean norm, and $\bm R, \bm R^* \in \R^2$ are independent random vectors with distribution $F$. Taking two independent samples $\{\bm r_i\}_{i = 1}^n$, $\{\bm r_j^*\}_{j = 1}^n$ of $F$, we estimate the ES \eqref{eq:es} by
\begin{equation} \label{eq:es estimate}
    \widehat{\mathrm{ES}}(F, \bm y) \coloneq \frac1{n} \sum_{i = 1}^{n} \lVert \bm r_i - \bm y \rVert - \frac1{2 n^2} \sum_{i = 1}^{n} \sum_{j = 1}^{n} \lVert \bm r_i - \bm r_j^* \rVert.
\end{equation}
Another proper scoring rule is the \textit{logarithmic score} ($\mathrm{LogS}$, \citealp{Good1952}), which is based on the negative logarithm of $f$ evaluated at $\bm y$, i.e. 
\begin{equation*}
    \mathrm{LogS}(F, \bm y) \coloneqq - \ln(f(\bm y)).
\end{equation*}
The \textit{variogram score} ($\mathrm{VS}$) of order $d > 0$ introduced by \cite{scheuerer2015variogram}, is a proper scoring rule which measures pairwise distances between the components of $\bm y$. Compared to other multivariate proper scoring rules, such as the ES, the variogram score is more sensitive to correlation misspecifications. We use the unweighted VS of order $d = 0.5$ recommended by \cite{scheuerer2015variogram} given by
\begin{equation} \label{eq:vs}
    \mathrm{VS}(F, \bm y) \coloneqq 2{\left(\sqrt{\lvert y_1 - y_2 \rvert} - \mathds{E}_F \sqrt{\lvert R_1 - R_2 \rvert}\right)}^2,
\end{equation}
where $(R_1, R_2) \in \R^2$ is a random vector with distribution $F$. Using an independent sample $\{(r_{i1}, r_{i2})\}_{i = 1}^{n}$ of $F$, we estimate the $\mathrm{VS}$ in \eqref{eq:vs} by
\begin{equation*}
    \widehat{\mathrm{VS}}(F, \bm y) \coloneq 2 {\left(\sqrt{\lvert y_1 - y_2 \rvert} - \frac1{n} \sum_{i = 1}^{n} \sqrt{\lvert r_{i1} - r_{i2} \rvert} \right)}^2.
\end{equation*}

For a scoring rule $S \in \{\mathrm{ES}, \mathrm{LogS}, \mathrm{VS}\}$ and a sequence of verifying observations $\bm y_{tl}, t = 1, \dots, T$ at a location $l = 1, \dots, L$ we report the mean score at location $l$ and the mean score aggregated over all locations as
\begin{equation}\label{eq: score loc mean}
    \overline{S}_{F, l} \coloneq \frac1{T} \sum_{t = 1}^T S(F_l, \bm y_{tl}) \quad \text{and} \quad \overline{S}_F \coloneq \frac1{L} \sum_{l = 1}^L \overline{S}_{F, l},
\end{equation}
where $F_l$ is the forecast distribution at location $l$. 
For measuring the relative improvement of a probabilistic forecast distribution $F_l$ over a reference forecast distribution $F_{\text{ref}, l}$ at location $l$, we calculate a \textit{skill score} \citep{Murphy1973} from the mean scores of model $F_l$ and $F_{\mathrm{ref}, l}$, i.e.
\begin{equation*}
    \mathcal{SS}_l := 1 - \overline{S}_{F, l} / \overline{S}_{F_{\mathrm{ref}, l}}
\end{equation*}
This skill score is defined for non-negative scores such as $\mathrm{ES}$ and $\mathrm{VS}$. We denote them by $\mathrm{ESS}$ and $\mathrm{VSS}$, respectively. 

\paragraph{Sharpness.}
The \textit{determinant sharpness} ($\text{DS}$, \citealp{gneiting2008assessing}) assumes that the forecast distribution $F$ is Gaussian. For the bivariate case it is defined as 
\begin{equation*}
    \mathrm{DS}(F)\coloneq {(\det \bm{\Sigma})}^{\frac{1}{4}},
\end{equation*}
where $\bm{\Sigma} \in \R^{2 \times 2}$ is the covariance matrix associated with the distribution function $F$, which is estimated by the empirical covariance based on an independent sample of $F$. Similar to \eqref{eq: score loc mean} location specific and overall means for the determinant sharpness can be defined.

\paragraph{Calibration.}
In order to assess the calibration of a multivariate forecast distribution $F$, \cite{knuppel2022score} propose to use so-called \textit{score based PIT histograms} which are obtained in two steps. First, a strictly proper scoring rule $S$ is used to reduce the dimension of $\bm y \in \R^2$ to one. Then, score based PIT values are obtained via
\begin{equation} \label{eq:score based pit}
    U_S(F, \bm y) \coloneqq \mathds{P}_F (S(F, \bm R) \leq S(F, \bm y)),
\end{equation}
where $\bm R \in \R^2$ is a random vector with distribution $F$. Then $U_S$ is uniformly distributed and the score based PIT histogram is the PIT histogram using the realization of $U_S$. We use the energy score defined in~\eqref{eq:es} as the dimension reducing score $S$ in our analysis. The univariate energy score PIT value from~\eqref{eq:score based pit} is then estimated by 
\begin{equation} \label{eq:energy score based pit}
    \hat u_{\mathrm{ES}}(F, \bm y) = \frac1{n} \sum_{i = 1}^{n} \mathds{1} \left( \frac1{n} \sum_{i = 1}^{n} \lVert \bm r_i - \bm r_i^* \rVert \leq \frac1{n} \sum_{i = 1}^{n} \lVert \bm r_i - \bm y \rVert \right).
\end{equation}
The quantities in~\eqref{eq:energy score based pit} are identical to the ones in~\eqref{eq:es estimate}. 
We draw PIT histograms of the $\hat{u}_{ES}$ with $17$ bins $\left[ \frac{m - 1}{17}, \frac{m}{17} \right)$, $m = 1, \dots, 17$. To measure how far a PIT histogram deviates from being uniform the we calculate the \textit{reliability index} $\mathrm{RI} \coloneq \sum_{m = 1}^{17} \left| f_m - 1/17 \right|$,
where, $f_m$ is the relative frequency of PIT values in bin $m$. A smaller $\mathrm{RI}$ indicates a histogram closer to uniformity.

\paragraph{Testing for statistical significance.}
To assess whether the difference in performance metrics between two competing models is significant, we use the \textit{Diebold-Mariano test} \citep{Diebold1995}. It is applied to the series of score values of all verification cases for the two competing models. The test is carried out individually at each station, using the adapted large-sample standard normal test statistic of \cite{Moeller2016}. The type I error is adjusted by the \textit{Benjamini-Hochberg procedure} \citep{Benjamini1995} to account for the multiple testing setup according to \cite{Wilks2016}.
For all tests, a significance level of $\alpha=0.05$ is set.

\section{Case Study} \label{sec:results}
This section presents a case study for the data introduced in Section \ref{sec:data}, and compares the predictive performance for the seven bivariate postprocessing models in Table \ref{tab:overview_models}.

\subsection{Overall Results} \label{sec:overall_results}

\paragraph{Verification metrics and calibration}

The overall performance in terms of verification scores is presented in Table \ref{tab:scores} for the seven different models aggregated over the 60 stations and all 944 test days. 
With respect to ES, VS and LogS, \textbf{BIV-EMOS-GB}, \textbf{BIV-DRN} and all three Y-vine PP models clearly improve over the benchmark \textbf{IND-EMOS} as well as \textbf{BIV-EMOS}. While there is not much difference between the Y-vine PP model variants and \textbf{BIV-EMOS-GB}, \textbf{BIV-DRN} shows the best performance with respect to all three scores. 
\textbf{BIV-EMOS-GB} yields the smallest sharpness in terms of DS, closely followed by \textbf{BIV-DRN} and the Y-vine PP models. All models were estimated on a single cluster node with 40 CPUs. For each model, we additionally record the total estimation time $T$ across all stations. Notably, \textbf{BIV-YV-G} exhibits the shortest estimation time, which is due to the model consisting only of Gaussian pair copulas.

\begin{table}
    \centering
    \resizebox{0.75\linewidth}{!}{
    \begin{tabular}{lcccccc}
    \toprule[1.1pt]
     \textbf{Model}  & $\overline{\mathrm{ES}}$ & $\overline{\mathrm{VS}}$ & $\overline{\mathrm{LogS}}$ & $\overline{\mathrm{DS}}$ & RI & $T$ (min)\\ \midrule
     \textbf{IND-EMOS} & 1.1862 & 0.4207 & 3.3836 & 1.3313 & 0.0588 &  118\\
     \textbf{BIV-EMOS} & 1.1858 & 0.4160 & 3.3741 & 1.3218  & 0.0505 & 121\\
     \textbf{BIV-EMOS-GB} & 1.1137 & 0.3827 & 3.2598 & 1.2322  & 0.0631 & 55\\
     \textbf{BIV-DRN} & \textbf{1.0813} & \textbf{0.3647} & \textbf{3.1933} & 1.2748 & 0.1357 & 28\\
     \textbf{BIV-YV-G} & 1.1185 & 0.3819 & 3.3240 & 1.2729  & 0.1290 & \textbf{1}\\
     \textbf{BIV-YV-P} & 1.1138 & 0.3820 & 3.2736 & 1.2528  & \textbf{0.0193} & 294\\
     \textbf{BIV-YV-ALL} & 1.1147 & 0.3829 & 3.2768 & 1.2485  &  0.0204 & 354\\ 
     \bottomrule[1.1pt]
    \end{tabular}
    }
    \caption{Estimated verification scores averaged over the 60 stations and 944 dates in the test data. For the Y-vine PP models the LogS was infinity for three test days, these cases were removed. Bold numbers represent the best value. An exception is the DS, where the value cannot be judged by itself, without taking calibration into account.}
    \label{tab:scores}
\end{table}

\begin{figure}
    \begin{center}
		\includegraphics[width = 0.5\linewidth]{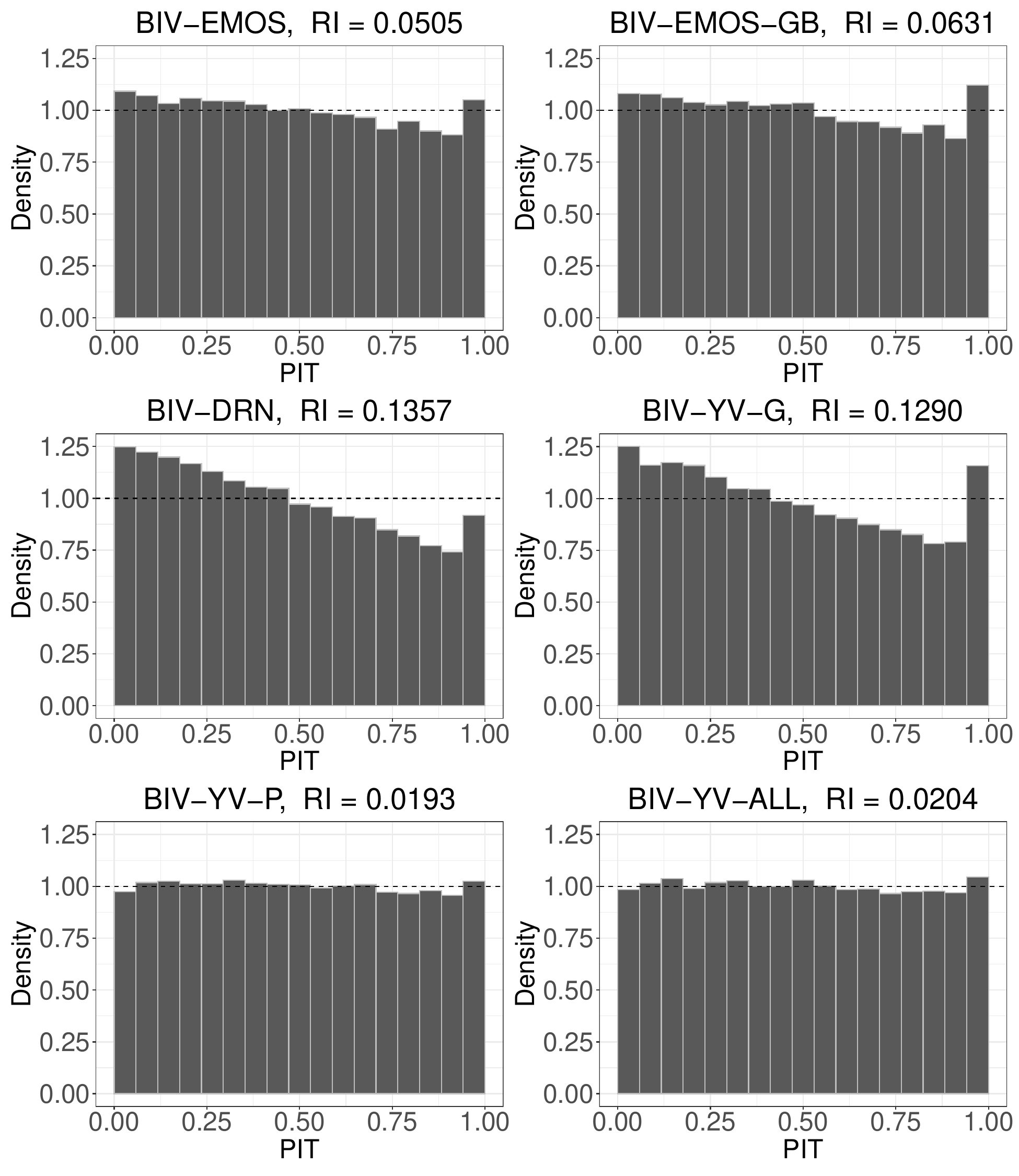}
    \end{center}	
    \caption{ES-based univariate PIT histograms aggregated over 60 stations and 944 days in the test data for six models. RI is the reliability index.}
    \label{fig: pit_all}
\end{figure}

Figure \ref{fig: pit_all} shows the ES-based PIT histogram for six models along with the reliability index (RI) computed from the given histograms. As \textbf{IND-EMOS} is a benchmark model not incorporating any bivariate dependence, it was omitted here. However, the RI of \textbf{IND-EMOS} can be found in Table \ref{tab:scores}, it is slightly higher than the RI of \textbf{BIV-EMOS}. 
The Y-vine PP model \textbf{BIV-YV-P} allowing all parametric copula families as well as the Y-vine PP model \textbf{BIV-YV-ALL} additionally allowing non-parametric copula families exhibit histograms closest to uniformity, where the RI of \textbf{BIV-YV-P} (0.0193) is even slightly lower than the RI of \textbf{BIV-YV-ALL} (0.0204). The Y-vine PP model consisting of only Gaussian copulas exhibits a pronounced bias indicated by a skewed histogram and an over occupied final bin, accompanied by a higher RI  (0.1290). This behavior is even stronger for \textbf{BIV-DRN}, coming along with a further increased RI (0.1357), while \textbf{BIV-EMOS} and \textbf{BIV-EMOS-GB} exhibit a similar level of calibration, both with a slightly over occupied final bin and similar RI. According to \cite{knuppel2022score} a decreasing shape of the score based PIT histogram indicates overconfident forecasts.

\begin{figure}
    \begin{center}
		\includegraphics[width=0.75\linewidth]{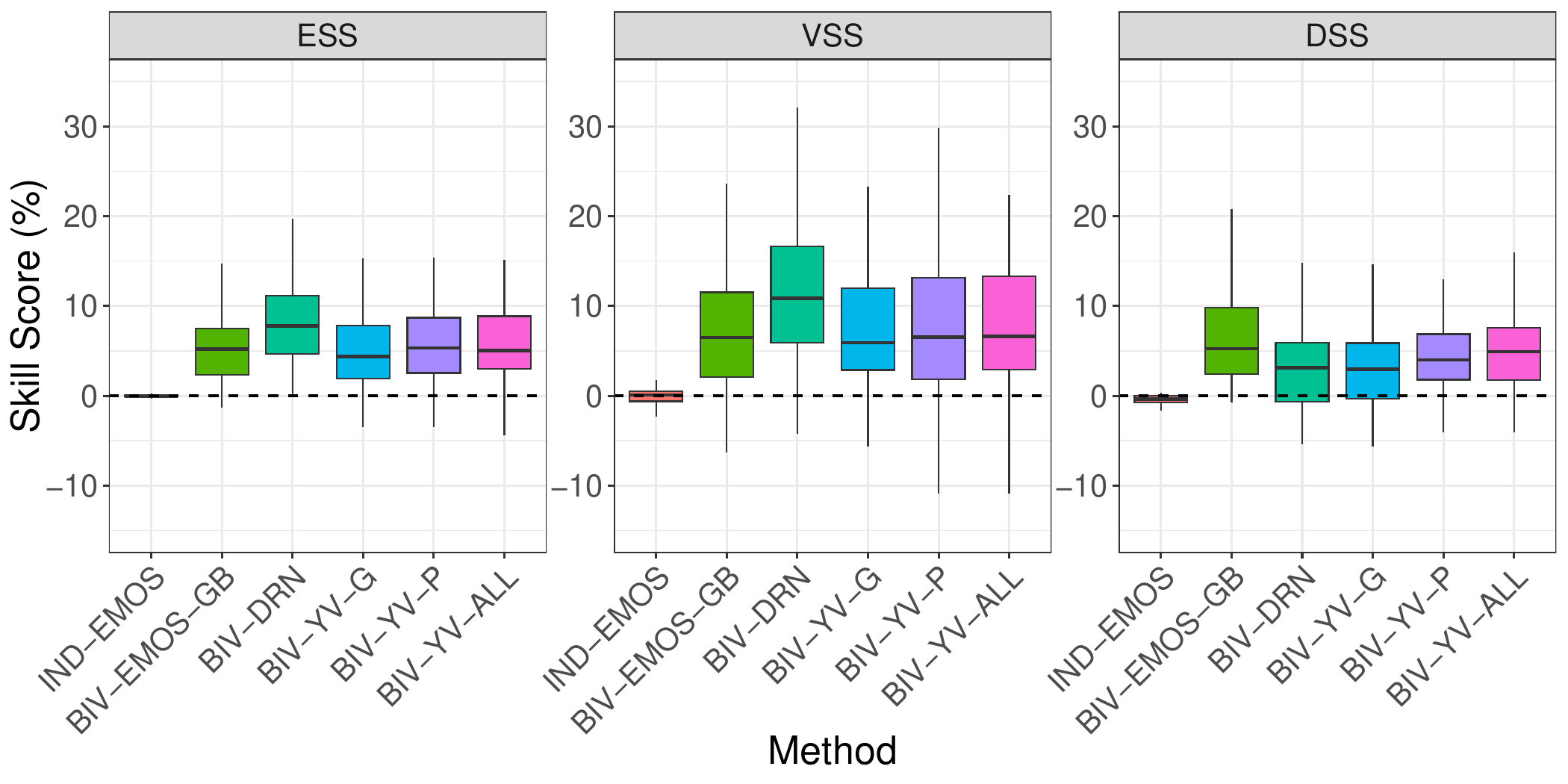}
    \end{center}	
    \caption{Boxplots without outliers of station-wise skill score improvements (\%) over \textbf{BIV-EMOS} as reference method aggregated over 944 test days.}
    \label{fig: boxplots_ss}
\end{figure}

Figure \ref{fig: boxplots_ss} shows boxplots of EES (left panel), VSS (middle panel), and DSS (right panel) for the test data with \textbf{BIV-EMOS} as reference model. The boxplots visualize the skill score values across the 60 stations, for six models relative to the reference model. \textbf{IND-EMOS} shows essentially no improvement over the reference model \textbf{BIV-EMOS}. As \textbf{BIV-EMOS} models the correlation between the wind vector components while \textbf{IND-EMOS} assumes independence between the components conditional on the covariates, it was to be expected that it cannot improve over a bivariate model with non-zero correlation. 

\begin{figure}
    \centering
    \includegraphics[width=0.5\linewidth]{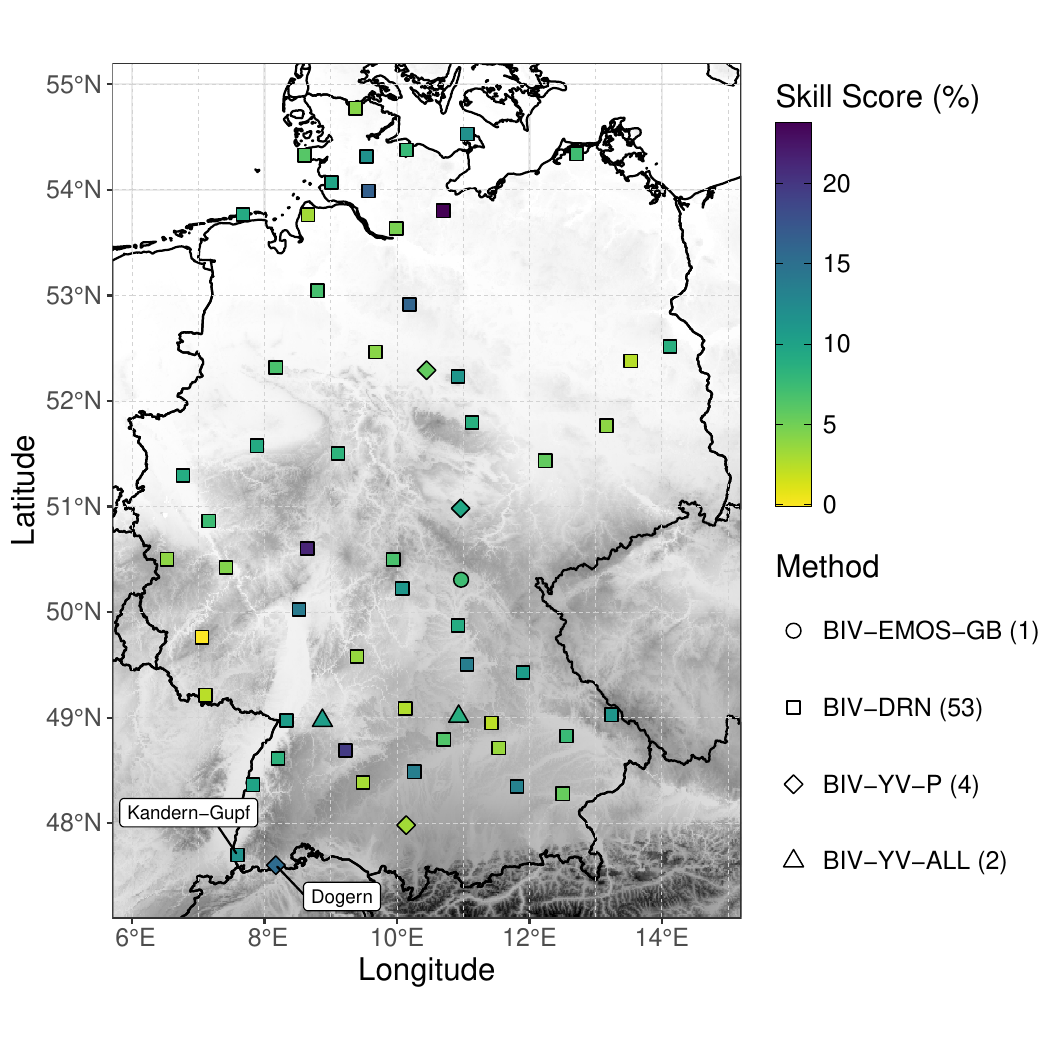}
    \caption{Method with the highest skill score improvement (\%) over reference method \textbf{BIV-EMOS} in terms of ES at each station in the test data.}
    \label{fig: ss_map}
\end{figure}

For both ESS and VSS, the largest improvements over \textbf{BIV-EMOS} are exhibited for \textbf{BIV-DRN}. While for $50 \%$ of the stations the improvement in terms of ESS is roughly between $5\%$ and $12\%$, there are a few stations for which the improvement is around $20 \%$. The improvements for the Y-vine PP models and \textbf{BIV-EMOS-GB} in terms of ESS are quite similar, showing improvements up to approximately $15 \%$. In general, the Y-vine PP models show a tendency to slightly higher improvements than \textbf{BIV-EMOS-GB}. Additionally, \textbf{BIV-VY-P} and \textbf{BIV-YV-ALL} show more cases with slightly higher improvements than \textbf{BIV-YV-G}, indicating that the use of non-Gaussian copula families leads to further improvements. With respect to VSS, the overall picture is similar. \textbf{BIV-DRN} shows the largest improvement over the reference model of roughly $5\%$ to $17 \%$ for $50 \%$ of the stations and admits improvements of more than $30 \%$ for individual stations. \textbf{BIV-EMOS-GB} and the three Y-vine PP models all show similar improvements over the reference model of around $5\%$ to $12 \%$. For DSS, the improvements in terms of sharpness with respect to the reference model are nearly similar for \textbf{BIV-DRN} and the three Y-vine PP models. We can observe improvements around $5\%$ to $8 \%$ for $50\%$ of the stations, with some stations showing improvements up to $15 \%$. Larger improvements are mainly found for \textbf{BIV-EMOS-GB}, which can be up to $20 \%$. However, sharpness is not of much relevance as an isolated characteristic, but is always relevant subject to the level of calibration of the predictive distribution. Therefore, the right panel showing the DSS cannot be interpreted as isolated information. Yet, we observe that for models showing (stronger) improvements in terms of ESS or VSS, indicated by a higher skill score, the improvement in sharpness seems always to go along.

\begin{figure}
    \begin{center}
		\includegraphics[width=0.5\linewidth]{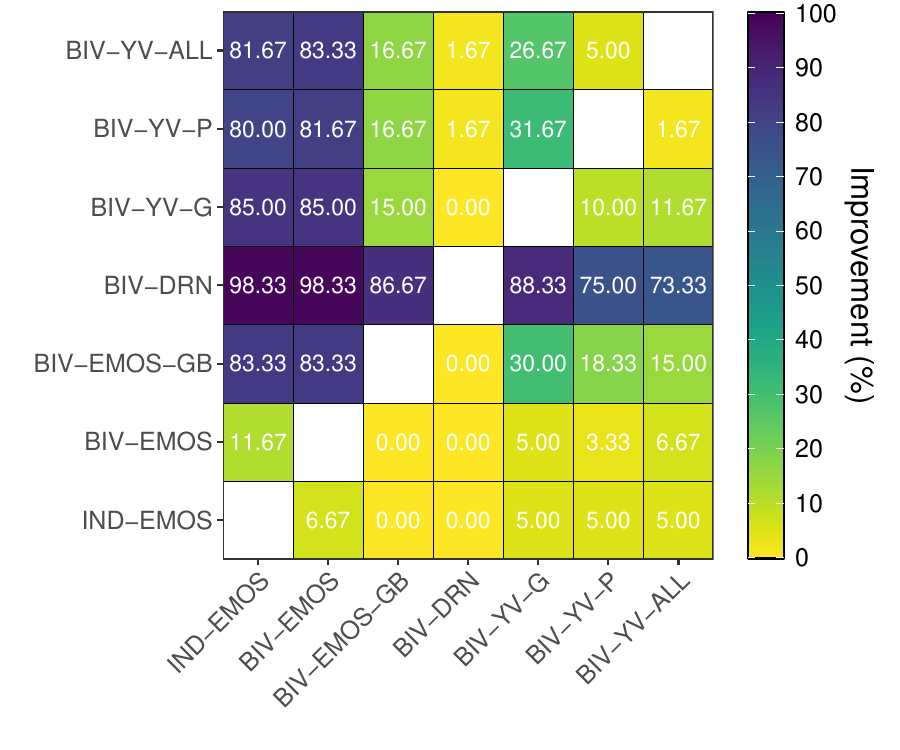}
    \end{center}	
    \caption{Percentage of stations for which the Diebold-Mariano (DM) tests indicate statistically significant ES improvements of the method in the row over the method in the column for a significance level of $\alpha = 0.05$. The $p$-values have been corrected for multiple testing using the Benjamini–Hochberg procedure.}
    \label{fig: dm_es}
\end{figure}

Figure \ref{fig: dm_es} shows the percentage of stations at which the Diebold-Mariano test indicated a statistically significant improvement in terms of ES for the method depicted in the row over the method depicted in the column. The color codes the percentage of stations with significant improvement. 
It is clearly visible that the three Y-vine PP methods and \textbf{BIV-EMOS-GB} improve over the two basic bivariate EMOS models at $80\%$ to $85 \%$ of the stations, while \textbf{BIV-DRN} shows an improvement at around $98 \%$ of the stations. \textbf{BIV-DRN} also improves over \textbf{BIV-EMOS-GB} at approx. $87 \%$ of the stations. 
Overall, \textbf{BIV-DRN} provides the largest improvement over \textbf{IND-EMOS} and \textbf{BIV-EMOS}, and it also shows substantial improvement over \textbf{BIV-EMOS-GB} and the Y-vine PP models. 

Figure \ref{fig: ss_map} shows the color-coded station-wise ESS of four models over the reference model \textbf{BIV-EMOS}. The point symbol represents the method with the highest skill score improvement at that station. The legend at the bottom right provides the number of stations in parentheses at which each of the considered models performs best in terms of improvement over the reference model. 
\textbf{BIV-DRN} shows the highest improvement for most of the stations (53 out of 60), while the other three considered models only perform best at a few of the stations. The achieved improvement can be up to approximately $20\%$ to $25 \%$, specifically for \textbf{BIV-DRN}. The largest improvements are geographically located in the center of Germany, along the north-south axis of the country. In the eastern and western part of the country, the improvements are slightly lower. \textbf{BIV-DRN} is also the method with the highest improvement over the reference model at Kandern-Gupf, considered in more detail in Section \ref{sec:station_results}. The improvement at that station is around $10\%$.

\begin{figure}
    \begin{center}
		\includegraphics[width = 0.75\linewidth]{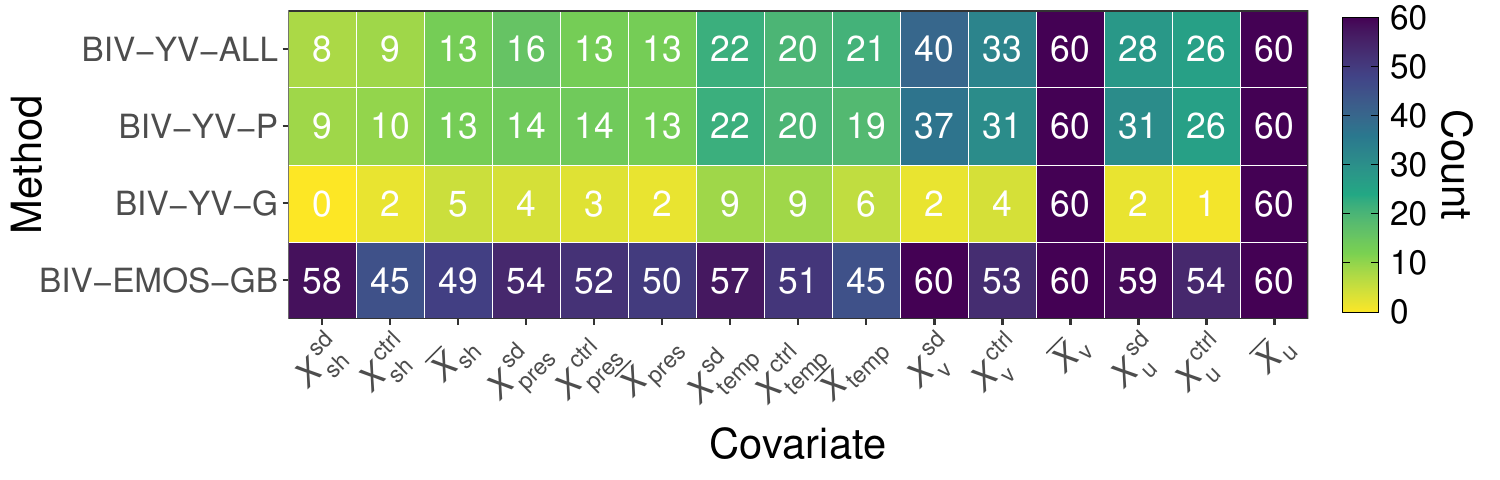}
    \end{center}	
    \caption{Counts of selected covariates across the 60 stations for 4 different models that are able to perform variable selection.}
    \label{fig: predictor_selection}
\end{figure}

\paragraph{Analysis of selected covariates.}
Figure \ref{fig: predictor_selection} displays the absolute frequency with which each covariate is selected across all stations in the models \textbf{BIV-EMOS-GB}, \textbf{BIV-YV-G}, \textbf{BIV-YV-P}, and \textbf{BIV-YV-ALL}. For a Y-vine PP model, a covariate is considered to be selected if it is included in the fitted Y-vine copula at a given station. In contrast, for \textbf{BIV-EMOS-GB} at a single station, a covariate is counted as selected if at least one of its regression coefficients across all five linear predictors is non-zero. In terms of the number of selected covariates, \textbf{BIV-EMOS-GB} is the most greedy and \textbf{BIV-YV-G} is the most parsimonious method. \textbf{BIV-YV-P} and \textbf{BIV-YV-ALL} perform comparable in terms of the number of selected covariates. The empirical ensemble mean forecasts $\overline{X}_{u}$ and $\overline{X}_{v}$ are most frequently selected across all methods indicating their high importance, followed by the empirical ensemble standard deviation and the control forecast of the $u$- and $v$-wind components as well as of surface temperature. However, covariates which are not obviously directly related to the $u$- and $v$-wind components, e.g.\ arising from surface pressure or the logarithmic specific humidity are selected less frequently. Interestingly, the empirical ensemble standard deviation is more important than the control forecast for nearly all weather quantities across all methods.

\paragraph{Analysis of chosen pair copula families.}
\textbf{BIV-YV-G}, \textbf{BIV-YV-P}, and \textbf{BIV-YV-All} consist of pair copulas that model the (conditional) dependence between two random variables. The copula family and thus type of dependence for each pair copula can be chosen among available families. \textbf{BIV-YV-G} consists of only Gaussian pair copulas. The resulting multivariate copula is in this case a multivariate Gaussian copula. \textbf{BIV-YV-P} and \textbf{BIV-YV-All} automatically select the best fitting pair copula family based on a BIC penalized pair copula log-likelihood. Possible families are the independence copula and parametric copula families available in \cite{rvinecopulib2025} for \textbf{BIV-YV-P} and additionally non-parametric pair copulas for \textbf{BIV-YV-All}. Among all pair copulas in fitted Y-vine copulas of \textbf{BIV-YV-P} for the 60 stations, $26.3\%$ are the independence copula ($25.4\%$ for \textbf{BIV-YV-All}) and $10.1\%$ are the Gaussian copula ($10.5\%$ for \textbf{BIV-YV-All}). This indicates that it might not be sufficient to assume a multivariate Gaussian dependence structure for the data, as also reflected in Figure~\ref{fig: pit_all}.

\subsection{Analysis of individual stations} \label{sec:station_results}
The stronger the conditional dependence between the two observed wind components, the more benefit is expected from a model that explicitly incorporates this conditional dependence. Furthermore, in the ideal case the magnitude of the dependence is similar in the training and test data. 
Therefore, we analyze the empirical dependence between the observed $u$ and $v$ components given the covariates in training and test data at the 60 individual stations more closely in the following. Subsequently, we present a performance analysis of a specific station exhibiting a stronger empirical partial correlation \citep{czado2019} than most of the other stations.

\begin{figure}
    \begin{center}
		\includegraphics[width = 0.65\linewidth]{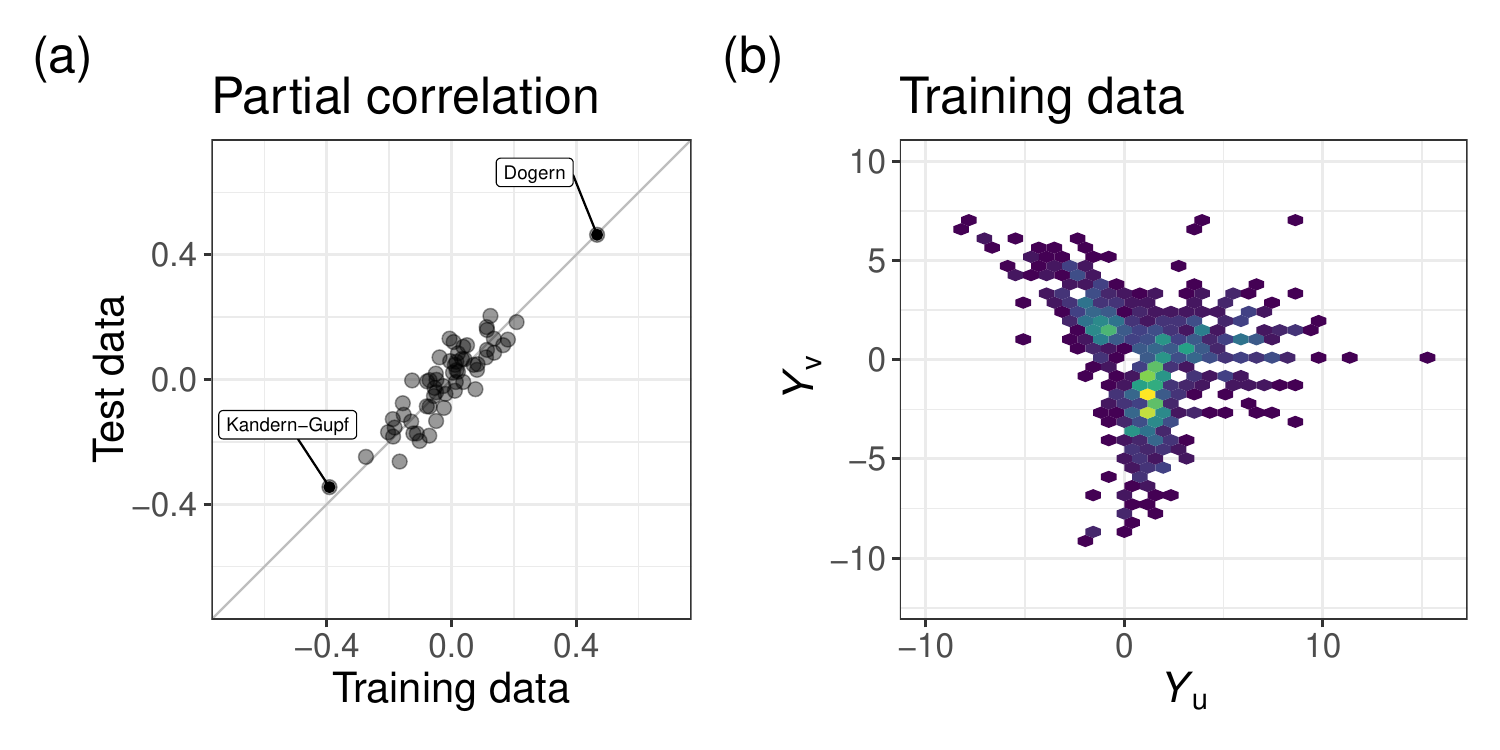}
    \end{center}	
    \caption{Empirical partial correlation for the observed $10$ m surface wind vector $u$- and $v$-component at each station for the training (a). Counts of the observed wind directions at Kandern-Gupf in the training data (b).}
    \label{fig: partialcorr}
\end{figure}

\paragraph{Initial dependence analysis.}
To quantify the conditional dependence between the responses, we estimate the empirical partial correlation between the observed $u$- and $v$-wind vector components at each station. A scatter plot of the empirical partial correlations in the training and test data for all stations is shown in Figure \ref{fig: partialcorr} (a).
The strongest partial correlation is observed at stations Dogern and Kandern-Gupf, which is much higher in absolute value than for the rest of the stations. In the following, we will exemplary investigate Kandern-Gupf in more detail. We also investigated Dogern in a preliminary analysis. As the results are similar to Kandern-Gupf, we will not present them here. 
Figure \ref{fig: partialcorr} (a) shows that the points corresponding to the stations are all relatively close to the bisecting line, indicating that the partial correlations in the training and test data are indeed similar for all stations. 
According to Figure \ref{fig: partialcorr} (b), 
Kandern-Gupf shows three dominant wind directions in the training data. All our methods, apart from \textbf{BIV-YV-ALL}, can model dependence structures which are elliptical or have at most two opposing dominant directions. This might not be sufficient for Kandern-Gupf as indicated in the left panel of Figure \ref{fig: enc_plot} and the inclusion of non-parametric pair copulas in \textbf{BIV-YV-ALL} could be beneficial, since they can model any type of dependence.

\begin{figure}[h!]
    \begin{center}
		\includegraphics[width = 0.60\linewidth]{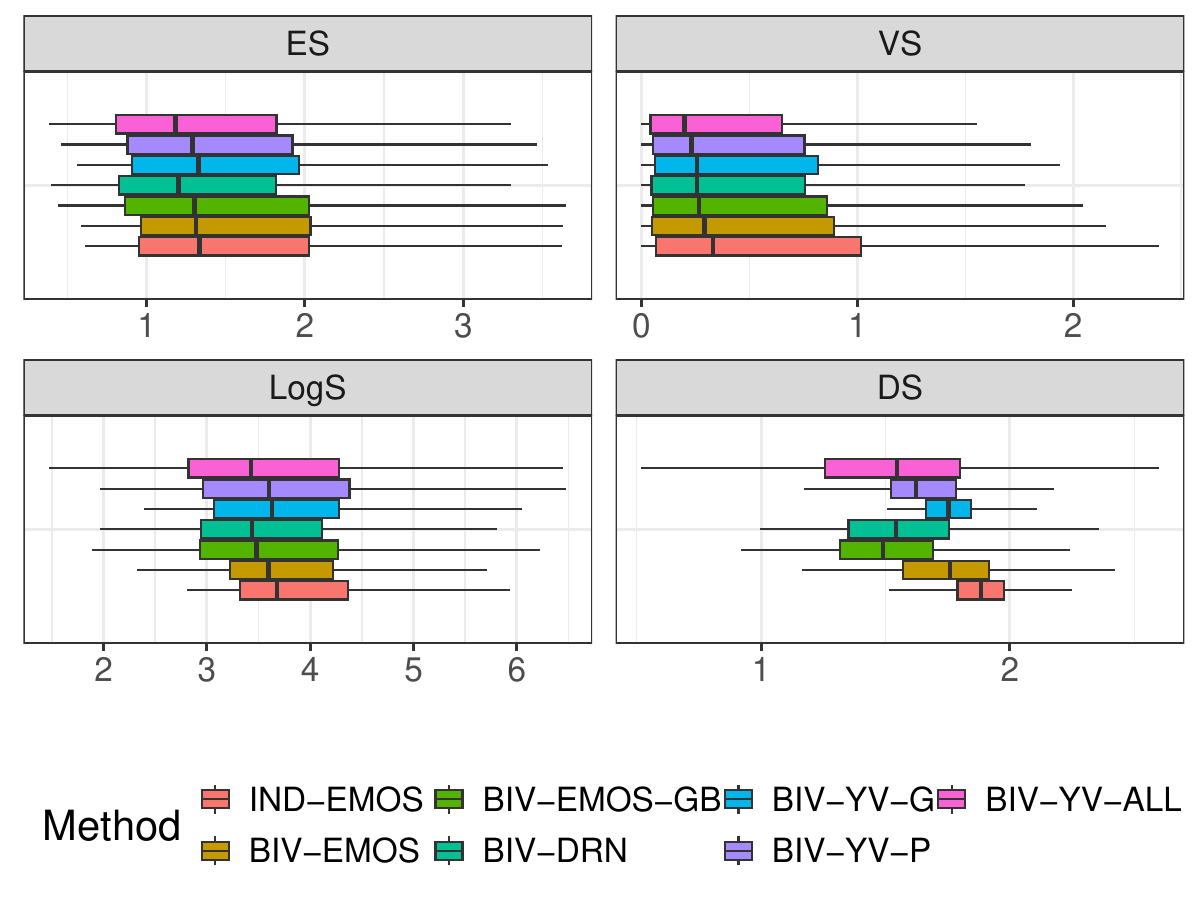}
    \end{center}	
    \caption{Boxplots without outliers for ES, VS, LogS and DS for all 944 days in the test data at Kandern-Gupf.}
    \label{fig:boxplots_kandern}
\end{figure}

\paragraph{Verification scores.}
Figure \ref{fig:boxplots_kandern} shows boxplots for ES, VS, LogS, and DS for all 944 test days for each of our models at Kandern-Gupf. The best performing model in terms of VS is \textbf{BIV-YV-ALL} allowing for parametric and non-parametric copula families. It is closely followed by \textbf{BIV-YV-P} and \textbf{BIV-DRN}. \textbf{BIV-YV-G} and \textbf{BIV-EMOS-GB} perform nearly similarly, while the two basic EMOS variants perform worst in terms of VS. 
The lowest ES was achieved by \textbf{BIV-YV-ALL} and \textbf{BIV-DRN}, while \textbf{BIV-YV-P} and \textbf{BIV-YV-G} have higher ES, followed by the remaining models. \textbf{BIV-YV-ALL} and \textbf{BIV-DRN} have a similar median LogS, but the LogS of the former model has a much wider range. Thus, the reliability of the forecast is more volatile compared to \textbf{BIV-DRN}. This can also be seen for the DS, where \textbf{BIV-YV-ALL} shows the sharpest but also vaguest forecasts. 
This large spread in sharpness could reflect the true uncertainty in the forecast well and therefore also explains \textbf{BIV-YV-ALL}'s good calibration in Figure \ref{fig: pit_all}.

\section{Conclusion and Outlook} \label{sec:outlook}
We provide a comparison of different bivariate models for joint postprocessing of the $u$- and $v$-wind vector components. The case study indicates that bivariate methods including dependence between responses, improve upon methods that ignore this dependence. Further, methods based on the bivariate Gaussian distribution show reduced calibration compared to methods where the resulting bivariate distribution is more flexible and potentially non-Gaussian. Overall, the bivariate neural network provides the best predictive performance at most stations. In our analysis, we ignored temporal and spatial aspects. Incorporating these into the models could improve performance further and allows for prediction at unobserved stations. How to reasonably integrate spatio-temporal information into the model is subject to further research. Combining the strength of distributional regression networks with vine copula-based models is also of interest.

\section*{Acknowledgments}
Annette M\"oller, Ferdinand Buchner and Claudia Czado acknowledge support by Deutsche Forschungsgemeinschaft (DFG) Grant 520017589; Annette Möller and David Jobst appreciate support by DFG Grant 395388010.

\section*{Data Availability Statement}
The observation data can be downloaded from the German Weather Service. The forecast data is available for scientific purposes upon request from the European Center for Medium Range Weather Forecasts.  



\bibliographystyle{plainnat}
\bibliography{references}

\end{document}